\documentclass[twocolumn]{article}

\usepackage{arxiv}
\usepackage{preamble}

\title{Data-efficient inverse design of spinodoid metamaterials}

\author{
	Max Rosenkranz\\
	Chair of Scientific Computing \\
    for	Systems Biology\\
	TU Dresden,
	01062 Dresden, Germany \\
	\And
	Markus K\"{a}stner\thanks{Corresponding author, email: \texttt{markus.kaestner@tu-dresden.de}.} \\
	Chair of Computational and\\
	Experimental Solid Mechanics\\
	TU Dresden, 
	01062 Dresden, Germany \\
	\And
	Ivo F. Sbalzarini \\
	Chair of Scientific Computing\\
    for Systems Biology\\
	TU Dresden, 
	01062 Dresden, Germany \\
}

\hypersetup{
pdftitle={Preprint_RosenkranzEtAl_2024},
pdfauthor={Max Rosenkranz},
}

\begin{document}

\twocolumn[
\maketitle
\begin{abstract}
We create an data-efficient and accurate surrogate model for structure-property linkages of spinodoid metamaterials with only 75 data points -- far fewer than the several thousands used in prior works -- and demonstrate its use in multi-objective inverse design.
The inverse problem of finding a material microstructure that leads to given bulk properties is of great interest in mechanics and materials science.
These inverse design tasks often require a large dataset, which can become unaffordable when considering material behavior that requires more expensive simulations or experiments.
We generate a data-efficient surrogate for the mapping between the characteristics of the local material structure and the effective elasticity tensor and use it to inversely design structures with multiple objectives simultaneously. The presented neural network-based surrogate model achieves its data efficiency by inherently satisfying certain requirements, such as equivariance with respect to permutations of structure parameters, which avoids having to learn them from data. The resulting surrogate of the forward model is differentiable, allowing its direct use in gradient-based optimization for the inverse design problem. We demonstrate in three inverse design tasks of varying complexity that this approach yields reliable results while requiring significantly less training data than previous approaches based on neural-network surrogates. This paves the way for inverse design involving nonlinear mechanical behavior, where data efficiency is currently the limiting factor.


\end{abstract}
]


\clearpage

\section{Introduction}
\label{intro}
\begin{figure*}
    \centering
    \includegraphics[width=\textwidth]{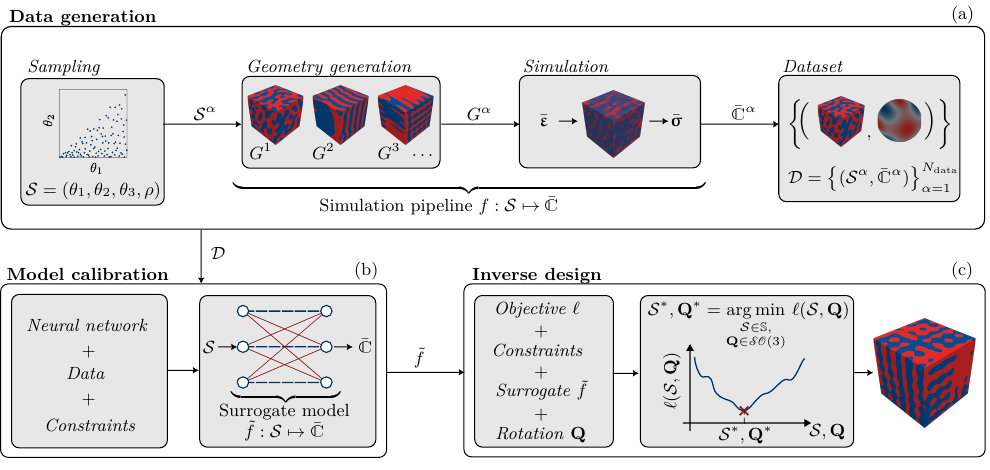}
    \caption{Overview of the workflow used here: (a) A dataset $\dataset$ consisting of structure parameters $\strpar^\alpha$ and corresponding effective elasticity tensors $\eltens^\alpha$ is generated. Points in the four-dimensional parameter space $\strparspc$ are sampled appropriately, the corresponding geometries $\geo^\alpha$ are created, and the effective elasticity tensor is determined computationally. Geometry generation and simulation are abbreviated as $\fwdmod : \strpar\mapsto\eltens$. (b) The generated dataset is used to calibrate the proposed surrogate model $\surmod$, which replaces $\fwdmod$. (c) Inverse design is formulated as the minimization of an objective function $\ell$ with respect to the structure parameters $\strpar$ and additionally rotations $\rottens$, which can be solved efficiently using the surrogate model.}
    \label{fig:Overview}
\end{figure*}

Materials with application-adapted behavior are of great interest in engineering applications, as they allow to further optimize the performance of a part as well as resource efficient production. Unlike conventional materials such as alloys, whose properties are dictated by composition and processing conditions, architected materials or \emph{metamaterials} introduce additional design freedom by relying not only on the base materials $\material_\alpha$ itself, but also using the structure $\structure$ on the mesoscale to control the effective properties $\prop$. Thus, a metamaterial $\metamaterial(\structure, \material_\alpha)$ is defined by both the materials it is made of as well as the mesostructure describing the spatial arrangement of those materials. 
The effective properties of interest can be, for example, the stiffness \cite{kumar_inverse-designed_2020,van_t_sant_inverse-designed_2023,yu_mechanical_2018,cheng_design_2022,berger_mechanical_2017,tancogne-dejean_3d_2018}, energy absorption characteristics \cite{yuan_3d-printed_2019,mohsenizadeh_additively-manufactured_2018}, or the type of anisotropy \cite{kadic_anisotropic_2013,zhu_optimisation_2021,jiang_elastically_2023,jadoon_inverse_2024}. Typical examples of extraordinary properties that can be achieved with architected materials are negative Poisson's ratios \cite{wang_novel_2020,cheng_design_2022,lakes_foam_1987} or even negative stiffnesses \cite{wang_negative_2004,mehreganian_structural_2021,liu_viscoelastic_2025,correa_negative_2015,correa_mechanical_2015}. Architected materials are therefore particularly interesting for the development of novel materials with application-specific properties.

Traditionally, material development has relied on empirical approaches and forward modeling, where a novel material is developed, fabricated, and tested iteratively to achieve the desired properties. This process, however, is resource-intensive, requiring large experimental efforts for both traditional materials \cite{vecchio_high-throughput_2021,yeh_alloy_2013,hariharan_highspeed_2025}, as well as for architected materials. Numerical experiments significantly reduce this effort. However, pure trial-and-error approaches for finding a structure with desired properties remain inefficient. Therefore, more sophisticated approaches for \emph{inverse design} have been developed in recent years.

In the most general setting, the problem of inverse designing metamaterials can be stated as follows: \emph{Find a metamaterial $\metamaterial(\structure, \material_\alpha)$ with mesostructure $\structure$ and base materials $\material_\alpha$, such that $\metamaterial$ has desired effective properties $\prop$.} Solving the inverse design problem in this generality is difficult. Therefore, the problem is restricted as follows within the scope of this work: (i) the base materials are fixed, thus only a suitable structure $\structure$ has to be identified, (ii) the structure $\structure$ belongs to a class of structures that can be described by means of structure parameters $\strpar$ and (iii) there is a mapping
\begin{equation}
    \label{eq:fwdmod}
    \fwdmod: \strpar \mapsto \prop
\end{equation}
available, which determines the effective properties for the metamaterial with structure resulting from given structure parameters $\strpar$. The mapping $\fwdmod$ is often termed the structure-property-relation. This could be an experiment or a simulation. An intuitive idea to solve this inverse problem is to find the inverse $\fwdmod^{-1}: \prop \mapsto \strpar$. However, as $\fwdmod$ is usually not bijective, this inverse does not exist.

A wide range of different approaches for inverse design exist, each with its own advantages and limitations.
Some methods use the forward process $\fwdmod$ as it is \cite{rasloff_inverse_2025}, some make use of a \emph{surrogate model} \cite{kumar_inverse-designed_2020}
\begin{equation}
    \surmod : \strpar \mapsto \prop\,,\quad\text{where}\quad \surmod(\strpar)\approx\fwdmod(\strpar) \quad .
\end{equation}
Methods without surrogate models rely on evaluating $\fwdmod$ and typically do not require large pre-existing datasets, but can still be expensive if evaluating $\fwdmod$ is expensive. On the other hand, surrogate-based inverse design utilizes pre-trained models $\surmod$ that approximate the mapping $\fwdmod$ between structure information and effective properties, providing an efficient and differentiable replacement to real or numerical experiments with short inference time. Once trained, a surrogate model enables several other inverse design techniques, that would not be feasible without surrogate.
Regardless of whether a surrogate model is used or not, the methods can be categorized into indirect inverse design and direct inverse design.

\emph{Indirect inverse design} predicts material properties using forward structure-property relations 
and then applies optimization or selection techniques to identify optimal structures. High-throughput screening, for example, searches large databases for structures with properties close to the target, commonly used in alloy design \cite{calderon_aflow_2015,jin_accelerated_2021,rittiruam_high-throughput_2022}. Bayesian optimization iteratively suggests new designs by autonomously exploring the design space and simultaneously minimizing an objective using Gaussian process regression \cite{rasloff_inverse_2025,hanaoka_bayesian_2021,kusampudi_inverse_2023,zhang_inverse_2024}. Evolutionary algorithms have also been used to improve designs~\cite{coli_inverse_2022}. Physics-augmented models are used for data-efficient learning of material behavior dependent on structure parameters \cite{thakolkaran_experiment-informed_2025}.

A common approach for \emph{direct inverse design} is the combination of forward and inverse models, where the purpose of the forward model is the more efficient training of an inverse model, that directly suggests geometries or structure parameters based on desired properties.
Kumar et al. \cite{kumar_inverse-designed_2020} use a forward surrogate trained on 19,170 structure-property pairs to train another neural network for the inverse map in order to design spinodoid metamaterials. Using a similar approach, Van 't Sant et al.~\cite{van_t_sant_inverse-designed_2023} inversely design growth-based architected materials using 800,000 data points.
Bastek et al. \cite{bastek_inverting_2022} leverage stochastic and physics-guided neural networks trained on millions of data points to design truss-based materials. 
Other methods such as video denoising diffusion models have been introduced to directly design two-dimensional cellular metamaterials with targeted stress-strain responses \cite{bastek_inverse_2023}. Generative Adversarial Networks (GANs) \cite{liu_generative_2024} and Cycle-GANs \cite{tian_machine_2022}, have been used for generating two-dimensional spinodoid and cellular structures. While direct inverse design methods provide models for efficient geometry generation or suggestion of structure parameters, they often require extensive training datasets, limiting their data efficiency.

A prerequisite for data-efficient inverse design is to identify classes of materials that balance the trade-off between low design space dimensionality and flexibility of properties. Data generation is costly, making large datasets impractical, especially when targeting complex mechanical behaviors beyond linear elasticity. Therefore, the optimal type of material should offer a broad property space while being sufficiently described within a low-dimensional design space for efficient exploration. 
Among various types of architected materials \cite{guo_inverse_2024,meyer_non-symmetric_2024,wiesenhuetter_triply_2023,maurizi_inverse_2022}, spinodoids \cite{kumar_inverse-designed_2020} stand out in this regard. 

\emph{Spinodoid metamaterials} have been successfully designed for linear elasticity \cite{kumar_inverse-designed_2020}, targeted stress-strain curves \cite{jin_characterization_2025,thakolkaran_experiment-informed_2025}, electric properties  \cite{shi_3d_2024}, diffusivity \cite{roding_inverse_2022} or sound absorption \cite{wojciechowski_additively_2023}.
When spinodoid structures are inversely designed for desired target properties, typically large datasets with several thousand pairs of geometries (or structure parameters) and corresponding properties are used to obtain precise results, and/or the design space is restricted. While the number of required samples is not a major concern for computationally inexpensive properties, such as linear elastic behavior, data-efficiency becomes more important for more complex properties like nonlinear elastic or inelastic behavior. In particular, indirect methods such as Bayesian optimization \cite{rasloff_inverse_2025} aim to reduce the number of required data points by avoiding the use of an expensive pre-trained surrogate model.

Here, we show how inverse design of spinodoid metamaterials can be made data-efficiently by using concepts from geometric learning in the neural-network surrogate model. We quantify the expected gains using the example of linear elastic properties. Specifically, we determine the smallest number of data points needed to train a sufficiently precise surrogate model for the elasticity tensor.
The results suggest that even computationally expensive properties for inelastic behavior, for example, can be modeled with a surrogate at a reasonable computational cost, which can then be used for arbitrary inverse design tasks.

The present workflow is shown in \reff{fig:Overview}. First, \refs{sec:spinodoids} provides an overview of how to generate spinodoid structures, followed by a specific neural network architecture in \refs{sec:penn}, which is used in the formulation of the surrogate model described in  \refs{sec:surrogate_model}. The inverse design framework is then presented in \refs{sec:inverse_design}. The datasets used for training the surrogate are generated according to the approach presented in \refs{sec:data_generation}. In \refs{sec:applications}, we analyze which dataset sizes yield sufficiently accurate surrogates for three inverse design tasks of varying complexity.

\paragraph{Notation} The space of tensors and matrices of rank $r$ in $d$ dimensions is denoted as $\tensspc{d}{r}=(\realnums^d)^{\dyad r}$. We use italic symbols for scalar quantities ($P\in\realnums$), bold italic symbols for rank one tensors / vectors ($\tensorone{P}\in\tensspc{d}{1}$), upright bold symbols for rank two tensors / matrices ($\tensortwo{P}\in\tensspc{d}{2}$), blackboard bold symbols for fourth order tensors ($\tensorfour{P}\in\tensspc{d}{4}$) and script style symbols for tensors of rank eight ($\tensoreight{P}\in\tensspc{d}{8}$). Some symbols will denote matrix/tensor-valued quantities of unspecified rank, which we will denote with upright bold symbols too ($\tensorarb{P}$). 
The identity tensor of rank two is $\unittensor = \kronecker_{ij}\basisvec_i\dyad\basisvec_j\in\tensspc{3}{2}$, where $\delta_{ij}$ is the Kronecker symbol, $\basisvec_i\in\tensspc{3}{1}$ is the $i$-th cartesian basis vector and $\dyad$ denotes the dyadic product. Furthermore, $\unittensorfour=\frac{1}{2}\left( \kronecker_{ik}\kronecker_{jl} + \kronecker_{il}\kronecker_{jk} \right)\basisvec_i\dyad\basisvec_j\dyad\basisvec_k\dyad\basisvec_l\in\tensspc{3}{4}$ is the symmetric fourth order identity tensor. The symbols $\cdot$, $\dd$ and $\dddd$ denote single, double and quadruple contraction of adjacent indices, respectively. Effective quantities on the mesoscale are indicated with a bar $\bar\circ$. 

\section{Spinodoid structures}
\label{sec:spinodoids}
\begin{figure*}
    \centering
    \includegraphics{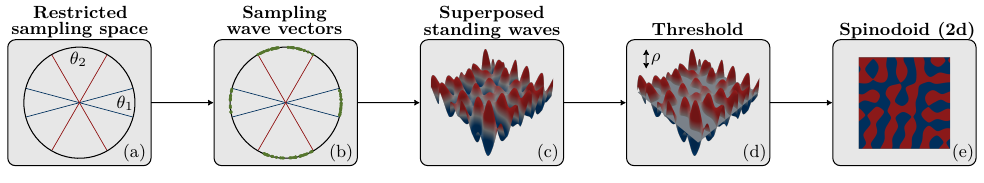}
    \caption{Generation of spinodoid structures using a 2d example: (a) The allowed sampling space of wave vectors is constrained by the half-angles $\mangle_i$. (b) A maximum number of cosine waves is sampled within the allowed region. (c) These cosine waves are superimposed to form a Gaussian random field. (d) Another parameter $\rho$ determines how much of the resulting structure is occupied by Material 1. Based on this parameter, the Gaussian random field is thresholded at a specific function value, and (e) everything below this threshold is interpreted as Material 1 (blue), while everything above is interpreted as Material 2 (red).}
    \label{fig:Spinodoid}
\end{figure*}

Spinodoid structures are a special class of metamaterials~\cite{kumar_inverse-designed_2020}, characterized in particular by their smooth geometrical features 
and versatile and adjustable properties despite their low-dimensional parameter space. The name ``spinodoid'' is derived from ``spinodal decomposition'', a specific form of phase separation in which a homogeneous initial phase splits into multiple distinct phases. The suffix ``-oid'' suggests that spinodoids are not exactly the topologies resulting from spinodal decomposition in a strict sense, but rather an approximation with similar-looking structures inspired by it. Spinodoids are generated by a Gaussian random field $\GRF$ of the form
\begin{equation}
    \label{eq:GRF}
    \GRF(\tensorone{x}) = \sqrt{\frac{2}{\Nwaves}}\sum_{i=1}^{\Nwaves} \cos(\wavenumber\wavevector_i\cdot\tensorone{x}+\phaseshift_i) \quad ,
\end{equation}
where $\tensorone{x}\in\realnums^3$ is the position in 3d space, $\wavevector_i\in\realnums^3$ with $|\wavevector_i|=1$ denotes the direction of the $i$th of $\Nwaves$ superimposed cosine waves and $\phaseshift_i$ is its phase angle. For a graphical representation of an equivalent 2d example, see \reff{fig:Spinodoid}. The directions $\wavevector_i$ are sampled uniformly from six (four in the 2d case) not necessarily connected subregions of the unit sphere. These subregions are determined by three angles (two for the 2d case) $\mangle_1, \mangle_2, \mangle_3$ that define the maximum allowed angles between the position vector of a point on the unit sphere $S^2$ and each coordinate axis. Formally, $\wavevector_i$ is sampled uniformly within
\begin{equation}
    \wavevector \sim \mathcal{U} \Big( \bigl\{ \matrixone{k}\in S^2 \mid \sumor\limits_{i\in \{1,2,3\}} \left(\left| \matrixone{k}\cdot\ \basisvec_i \right| > \cos{\mangle_i} \right) \bigl\} \Big)
\end{equation}
and $\phaseshift_i$ is sampled uniformly from $[0,2\pi)$.
Based on the Gaussian random field $\GRF$ described in \refe{eq:GRF}, a two-phase spinodoid metamaterial is defined using the indicator function
\begin{equation}
    \label{eq:indicatorfunc}
    \delta(\tensorone{x})=
        \begin{cases}
            0\,,\quad \text{if } \GRF(\tensorone{x}) \leq \GRF_0 \\
            1\,,\quad \text{if } \GRF(\tensorone{x}) > \GRF_0 \quad ,
        \end{cases}
\end{equation}
where $\GRF_0=\sqrt{2}\erf^{-1}(2\rho-1)$ splits the domain into two parts, such that the portion of points with $\delta(\tensorone{x})=0$ equals the specified volume fraction $\rho$.
All points with $\delta(\tensorone{x})=0$ are assigned material $M_1$ and all other points with $\delta(\tensorone{x})=1$ are assigned another material $M_2$.

In total, there are four parameters, that influence the morphology of spinodoids proposed in~\cite{kumar_inverse-designed_2020}: three angles $\theta_i$ restricting the sampling space of the wave vectors for the Gaussian random field and a volume fraction $\rho$ of material $M_1$. In order to guarantee connectivity of the resulting structure, these parameters are restricted to
\begin{align}
    \label{eq:strpardomain}
    \theta_i\in\manglespc \quad &\text{, where} \quad \manglespc=\left\{ \SI{0}{\degree}\right\}\cup(\SI{15}{\degree},\SI{90}{\degree}) \quad \text{and} \\
    \rho\in\rhospc \quad &\text{, where} \quad \rhospc = (0.3,1) \quad . \nonumber
\end{align}
Since $\manglespc$ is divided in two not connected regions, three types of spinodoid metamaterials are distinguished: \emph{lamellar} ones with only a single $\mangle_i$ different from zero, \emph{columnar} ones with two $\mangle_i$ being non-zero and \emph{cubic} spinodoids with no zero-valued morphological angles.
In the following, the four parameters are summarized in the tuple of structure parameters $\strpar=(\mangle_1, \mangle_2, \mangle_3, \rho)$ and the allowed domain defined in \refe{eq:strpardomain} is denoted as $\strparspc = \manglespc^3\times\rhospc$.
This description allows to design versatile structures with diverse properties using only four parameters. This simplicity in combination with the large range of properties makes spinodoid metamaterials particularly suitable for inverse design problems.

An interesting property of the spinodoid structures in the form presented here is the symmetry under permutations of the angles $\mangle_1, \mangle_2, \mangle_3$. Consider, for example, three parameter sets $\strpar_1=(15^\circ,0,0,0.5)$, $\strpar_2=(0,15^\circ,0,0.5)$ and $\strpar_3=(0,0,15^\circ,0.5)$. These three structures differ only in their preferred direction and in slight fluctuations due to random sampling. This observation is not limited to this specific parameter set, but generally applies to all permutations of the angles $\mangle_1, \mangle_2$ and $\mangle_3$. Consequently, these symmetries must also be reflected in the effective properties of the metamaterial, for example, in the effective elasticity tensor $\eltens$. Here, a specific permutation of the angles results in a renumbering of the four indices.
This insight means that it is not necessary to investigate the entire admissible parameter space but only a subdomain with $\mangle_1\geq\mangle_2\geq\mangle_3$, for example. All other information can then be derived from the information in that subdomain. This property will be exploited in the following to learn a function that covers the entire parameter space with as few data points as possible. For this purpose, a network architecture will be explained that strictly enforces this permutation equivariance without having to learn each of the six permutations independently.

\section{Permutation equivariant neural networks}
\label{sec:penn}
\begin{figure*}[t]
    \def\svgwidth{\linewidth}
    \includegraphics[width=\linewidth]{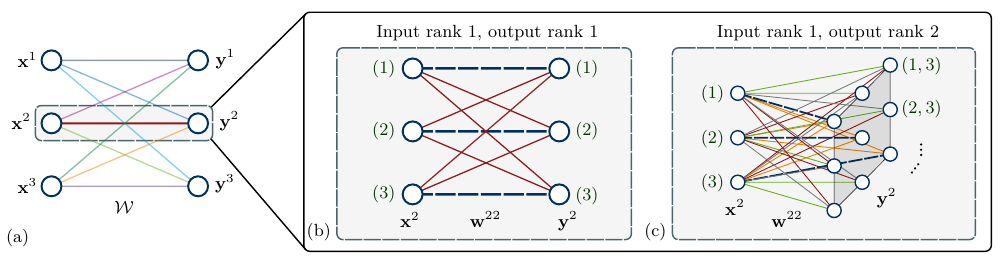}
    \caption{(a) A permutation equivariant layer, that maps three first order matrices $\tex^i$, i.e., $\rankx=1$, to three matrices of (b) rank one, i.e., $\ranky=1$ or (c) rank two, i.e., $\ranky=2$. The weights and biases of the layer are summarized in $\WeightsAndBiases$. Each connection $(i,j)$ is independent of the others, but is restricted internally to enforce the equivariance condition. More precisely, the weight matrix $\weightMatrix^{22}$ is constructed such that every orbit of input-output index pairs $\indexsetInput\times\indexsetOutput$ receives an independent weight. In case of rank one output matrices (b), this results in only two independent weights, one for horizontal connections and one for the diagonal ones. For a rank two output matrix (c), there are five weights. The same idea applies to any ranks $\rankx$ and $\ranky$.}
    \label{fig:PENN}
\end{figure*}

We are interested in functions whose outputs transform predictably under certain transformations of their inputs, i.e., they are equivariant under these transformations. 
Generally, a function $\phi: X\to Y,\,x\mapsto y=\phi(x)$ is said to be $\groupG$-equivariant, if
\begin{equation}
    \label{eq:defpermeq}
    \phi(\groupelement \actson x) = \groupelement\actson\phi(x) \quad \forall \groupelement\in\groupG, \forall x\in X\quad,
\end{equation}
where $\actson$ denotes the action of a permutation $\groupelement\in\groupG$ on $x\in X$ or $y\in Y$.
For the application in this contribution, we want to construct a parametrized function
\begin{align}
    &\phi_{\WeightsAndBiases}: (\tensspc{d}{\rankx})^{\Nin} \to (\tensspc{d}{\ranky})^{\Nout},\nonumber\\
    &(\tex^1,\ldots,\tex^{\Nin})\mapsto(\tey^1,\ldots,\tey^{\Nout})
\end{align}
of this type as a neural network layer with $\Nin$ inputs, $\Nout$ outputs and parameters $\WeightsAndBiases$. Herein, $\groupG$ will be the symmetric group $\symmetricgroup{d}$ with $d=3$, consisting of all $d!$ possible permutations of $d$ objects. That is, $\phi_{\WeightsAndBiases}$ has to be constructed such that any permutation $\groupelement$ acting on the inputs $\tex^i$ results in $\groupelement$-action on the outputs $\tey^j$.

As in a usual dense layer, each input node $\tex^i$ is connected with every output node $\tey^j$ via a weight $\weightMatrix^{ij}$ and every output has a bias $\biasMatrix^j$, such that
    \begin{equation}
        \label{eq:NN}
        \tey^j = \actfunc\left(\weightMatrix^{ij}\tex^i + \biasMatrix^j \right) \quad ,
    \end{equation}
where the summation convention is used and $\actfunc$ is an activation function acting on each coordinate independently. Consequently, the parameter set $\WeightsAndBiases = \bigl\{ \weightMatrix^{11}, \weightMatrix^{12}, \ldots, \allowbreak \weightMatrix^{\Nin\Nout}, \biasMatrix^1,\ldots,\biasMatrix^{\Nout} \bigl\}$ contains weights for every possible connection  between inputs and outputs and biases for every output. The major difference to standard dense layer is, that both $\tex^i$ and $\tey^j$ are matrices of arbitrary ranks \footnote{In this context, the term "rank" describes the number of indices of a matrix.} $\rankx$ and $\ranky$. Consequently, the weights $\weightMatrix^{ij}$ and biases $\biasMatrix^j$ are not scalars anymore, but matrices of rank $\rankx+\ranky$ and $\ranky$, respectively.

Since arbitrary values for $\weightMatrix^{ij}$ and $\biasMatrix^j$ will not yield a permutation equivariant network, the core idea presented in~\cite{ravanbakhsh_equivariance_2017} is to restrict the entries of $\weightMatrix^{ij}$ and $\biasMatrix^j$ via parameter sharing according to some rules.

Before explaining these rules, we first want to clarify how \refe{eq:defpermeq} with matrix-valued $x$ and $y$ is to be understood and what is meant by "permutation" when speaking of a matrix $\tex$. The matrix $\tex$ of rank $\rankx$ in $d$ dimensions has $d^{\rankx}$ entries at the index tuples ${}^\alpha \indexsetInputElement=({}^\alpha i_1,\ldots,{}^\alpha i_{\rankx})$. These possible index tuples of $\tex$ are summarized in $\indexsetInput=\{{}^\alpha \indexsetInputElement\}_{\alpha=1}^{d^{\rankx}}$. Likewise for the output matrix $\tey$, a single index tuple is ${}^\alpha \indexsetOutputElement=({}^\alpha j_1,\ldots,{}^\alpha j_{\ranky})$ and the set of all possible output index tuples is denoted as $\indexsetOutput=\{{}^\alpha \indexsetOutputElement\}_{\alpha=1}^{d^{\ranky}}$. With that, the action of $\groupelement\in\groupG$ on $\tex$ is herein defined as $\groupelement$ acting on all indices of $\tex$, i.e., $(\groupelement \actson \tex)_{\indexsetInputElement} = x_{\groupelement i_1,\ldots,\groupelement i_{\rankx}} = x_{\groupelement \cdot \indexsetInputElement}$. The same applies to the function value $\tey$ where $(\groupelement \actson \tey)_{\indexsetOutputElement} = y_{\groupelement j_1,\ldots,\groupelement j_{\rankx}} = y_{\groupelement \cdot \indexsetOutputElement}$. 

The approach of~\cite{ravanbakhsh_equivariance_2017} to construct $\weightMatrix$ can be summarized as follows: As a sum (like in $\weightMatrix^{ij}\tex^i$) of equivariant functions is again an equivariant function, all $\weightMatrix^{ij}$ and $\biasMatrix^{j}$ can be treated independently and we will write $\weightMatrix$ and $\biasMatrix$ for convenience. Two coordinate connections $(\indexsetInputElement,\indexsetOutputElement)\in\indexsetInput\times\indexsetOutput$ and $(\indexsetInputElement^{\prime},\indexsetOutputElement^{\prime})\in\indexsetInput\times\indexsetOutput$ share the same weight, if $(\indexsetInputElement,\indexsetOutputElement)$ and $(\indexsetInputElement^{\prime},\indexsetOutputElement^{\prime})$ share the same orbit under $\groupG$-action.
We denote the orbit of $(\indexsetInputElement,\indexsetOutputElement)$ as
\begin{equation}
    \groupG \circ (\indexsetInputElement,\indexsetOutputElement) = \left\{ (\groupelement\cdot\indexsetInputElement, \groupelement\cdot\indexsetOutputElement) \mid \groupelement\in\groupG \right\} \quad .
\end{equation}
I.e., if there exists a $\groupelement\in\groupG$, such that $(\groupelement\cdot\indexsetInputElement,\groupelement\cdot\indexsetOutputElement)=(\groupelement\cdot\indexsetInputElement^{\prime},\groupelement\cdot\indexsetOutputElement^{\prime})$, then $w_{\indexsetInputElement\indexsetOutputElement} = w_{\indexsetInputElement^{\prime}\indexsetOutputElement^{\prime}}$. In order to assemble $\weightMatrix$ accordingly, the set of pairs of input indices and output indices $\indexsetInput\times\indexsetOutput$ is partitioned into orbits under $\groupG$-action. Each orbit corresponds to a single, independent scalar valued weight and all entries of $\weightMatrix$ belonging to this orbit receive this exact same weight.
Similarly, the bias matrix $\biasMatrix$ is constructed such that every orbit of the output indices receives an independent value.

\paragraph{Example 1} Consider a layer with $d=3$, $r_{\tex}=1$, $r_{\tey}=1$ and the group $\groupG=\symmetricgroup{3}$ being the the symmetric group, as illustrated in \reff{fig:PENN}(b).
Input $\tex$ and output $\tey$ both share the same set of possible (one-valued) index tuples $\indexsetInput=\indexsetOutput=\left\{\idxtuple{1},\idxtuple{2},\idxtuple{3}\right\}$.
In order to find the symmetries of $\weightMatrix$ such that the resulting layer is $\groupG$-equivariant, $\indexsetInput\times\indexsetOutput$ is partitioned into orbits under $\groupG$-action.
For the edge $\idxtuplepair{1}{1}$, the orbit contains the three elements $\groupG\circ\idxtuplepair{1}{1}=\bigl\{ \idxtuplepair{1}{1}, \idxtuplepair{2}{2}, \idxtuplepair{3}{3} \bigr\}$.
Consequently, all edges in $\groupG\circ\idxtuplepair{1}{1}$ must receive the same weight indicated with the blue dashed lines in \reff{fig:PENN}(b).
There are six edges, that do not have a color yet, e.g., the edge $\idxtuplepair{1}{2}$.
The orbit $\groupG\circ\idxtuplepair{1}{2}$ of this edge 
contains all of the remaining edges.
Again, all edges in $\groupG\circ\idxtuplepair{1}{2}$ receive the same color which can be different from the color assigned to the other orbit $\groupG\circ\idxtuplepair{1}{1}$.
Since $\groupG\circ\idxtuplepair{1}{1}\cup\groupG\circ\idxtuplepair{1}{2}=\indexsetInput\times\indexsetOutput$, there cannot exist another orbit.
Thus, there are only two distinct orbits and consequently two independent weights, indicated with two different colors.

\paragraph{Example 2} A layer with $d=3$, $r_{\tex}=1$ and $r_{\tey}=2$ is to be constructed such that it behaves equivariant with respect to action of $\groupG=\symmetricgroup{3}$, see \reff{fig:PENN}(c). As in example 1, the set of index tuples for the input is $\indexsetInput=\left\{\idxtuple{1},\idxtuple{2},\idxtuple{3}\right\}$. But since $r_{\tey}=2$, $\indexsetOutput=\bigr\{\idxtuple{1,1}, \idxtuple{1,2}, \ldots, \idxtuple{3,3}\bigr\}$ contains nine elements. Partitioning $\indexsetInput\times\indexsetOutput$ into orbits yields five distinct orbits with representatives $\idxtuplepair{1}{1,1}$, $\idxtuplepair{1}{1,2}$, $\idxtuplepair{1}{2,1}$, $\idxtuplepair{1}{2,2}$ and $\idxtuplepair{1}{2,3}$. Each orbit receives another color, resulting in the graph in \reff{fig:PENN}(c).

\section{Surrogate model}
\label{sec:surrogate_model}
\begin{figure*}
    \centering
    \includegraphics{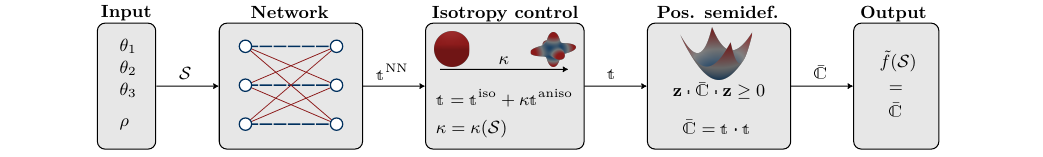}
    \caption{Architecture of the surrogate model $\surmod$: (a) The four structure parameters $\strpar$ are the input of $\surmod$. (b) A neural network, that is permutation equivariant w.r.t. $\theta_i$ is employed. This network additionally enforces the othorhombic symmetry of the output $\mathbbm{t}^{\text{NN}}$, as well as minor and major symmetry. (c) To ensure, that the output is isotropic if $\rho=1$ or any $\mangle_i=\SI{90}{\degree}$, the ansisotropic part is filtered out if needed and $\mathbbm{t}$ is obtained. (d) Squaring $\mathbbm{t}$ ensures positive semidefiniteness and (e) the output is interpreted as elasticty tensor $\eltens$.}
    \label{fig:Surmod}
\end{figure*}
The function $\fwdmod$ as defined in \refe{eq:fwdmod} takes structure parameters $\strpar = (\mangle_1, \mangle_2, \allowbreak \mangle_3, \rho)$ and calculates the effective elasticity tensor $\prop = \eltens$ for the corresponding structure. This function $\fwdmod$ can be a numerical or real experiment. Since the evaluation of $\fwdmod$ for a given $\strpar$ is expensive and $\fwdmod$ is in general not differentiable, it is favorable to create a differentiable surrogate model $\surmod: \strpar \mapsto \eltens$ with short inference time, that approximates $\fwdmod$ with reasonable accuracy.

There are a few requirements on the model that should be met to obtain a realistic model. The following are to be incorporated into the architecture presented here:
\begin{enumerate}[label=(\roman*), leftmargin=*]
    \item $\surmod$ ensures equivariance with respect to permutations of $\mangle_1, \mangle_2, \mangle_3$,
    \item the output $\eltens$ has minor and major symmetry, i.e., $\bar C_{ijkl}=\bar C_{jikl}$ and $\bar C_{ijkl}=\bar C_{klij}$,
    \item $\eltens$ reflects the (approximate) orthorhombic symmetries of the structure,
    \item $\eltens$ is isotropic, if $\rho=1$ or $\theta_i=\SI{90}{\degree}$ for any $i$, and
    \item $\eltens$ is positive semidefinite.
\end{enumerate}
As an ansatz for $\surmod$, a learnable function in the form of a neural network with parameters $\WeightsAndBiases$ is used. The five requirements are incorporated with the following techniques:
\begin{enumerate}[label=(\roman*), leftmargin=*]
    \item To guarantee permutation equivariance, a permutation equivariant neural network as described in \refs{sec:penn} is employed, where the inputs are a vector $(\mangle_1, \mangle_2, \mangle_3)$ and the volume fraction $\rho$ as matrix of rank zero.
    \item This only affects the last layer of the network. In order to guarantee minor and major symmetry, all orbits that contain the same elements after symmetrization of the output indices are merged. That is, e.g., the orbits of $(\indexsetInputElement,\indexsetOutputElement)=\idxtuplepair{1}{1,1,2,3}$ and $(\indexsetInputElement^\prime,\indexsetOutputElement^\prime)=\idxtuplepair{1}{2,3,1,1}$ receive the same weight.
    \item Requirement (iii) also only affects the final layer. It is utilized that the spinodoid structures used here are always aligned along the coordinate axes, i.e., the elasticity tensor, assuming orthorhombic symmetries, can be written as
        \begin{equation}
            \left[ \eltens \right] =
            \left[
            \begin{array}{@{\hskip 1pt}c@{\hskip 1pt}c@{\hskip 1pt}c@{\hskip 1pt}c@{\hskip 1pt}c@{\hskip 1pt}c@{\hskip 1pt}}
            \bar C_{1111} & \bar C_{1122} & \bar C_{1133} &        0 &        0 &        0 \\
            \bar C_{1122} & \bar C_{2222} & \bar C_{2233} &        0 &        0 &        0 \\
            \bar C_{1133} & \bar C_{2233} & \bar C_{3333} &        0 &        0 &        0 \\
                   0 &        0 &        0 & \bar C_{2323} &        0 &        0 \\
                   0 &        0 &        0 &        0 & \bar C_{1313} &        0 \\
                   0 &        0 &        0 &        0 &        0 & \bar C_{1212} \\
            \end{array}
            \right]
        \end{equation}
        in matrix notation with zeros in the specified positions. These zeros correspond to the orbits of the output index tuples $\idxtuple{1,1,1,2}$, $\idxtuple{1,1,2,3}$, and $\idxtuple{1,2,1,3}$. No weights and biases are assigned to these orbits.
        \item  Enforcing isotropy for $\rho=1$ or $\mangle_i=\SI{90}{\degree}$ is done by decomposing the network output $\mathbbm{t}^{\text{NN}}$ additively into an isotropic part $\mathbbm{t}^{\text{iso}}$ and an anisotropic part $\mathbbm{t}^{\text{aniso}}$ and reducing the influence of $\mathbbm{t}^{\text{aniso}}$ for increasing $\rho$ and $\mangle$. The updated matrix $\mathbbm{t}$ is calculated as
        \begin{equation}
            \label{eq:IsoSplit}
            \mathbbm{t} = \mathbbm{t}^{\text{iso}} + \kappa(\strpar) \mathbbm{t}^{\text{aniso}} \quad ,
        \end{equation}
        where the anisotropy factor $\kappa(\strpar)\in(0,1)$ is defined such that $\kappa(\strpar)=0$, if $\rho=1$ or $\mangle=\SI{90}{\degree}$ and $\kappa(\strpar)>0$ otherwise, using the relation
        \begin{equation}
            \kappa(\strpar) = (1-\rho) \prod_{i=1}^{3}(1-\hat{\theta}_i) \quad ,
        \end{equation}
        where $\hat{\mangle}_i=\mangle_i / \SI{90}{\degree}$.
        The isotropic part of $\mathbbm{t}^{\text{NN}}$ is obtained using the projection $\mathbbm{t}^{\text{iso}}=\Piso \dddd \mathbbm{t}^{\text{NN}}$ with the eighth order isotropic projection tensor $\Piso = \POne\dyad\POne + \frac{1}{5}\PTwo\dyad\PTwo$, where $\POne = \frac{1}{3}\unittensor\dyad\unittensor$ and $\PTwo=\unittensorfour - \POne$. The anisotropic part is $\mathbbm{t}^{\text{aniso}} = \mathbbm{t}^{\text{NN}} - \mathbbm{t}^{\text{iso}}$.
        \item Finally, positive semidefiniteness is achieved by squaring the fourth order tensor, i.e.,
        \begin{equation}
            \eltens = \mathbbm{t}\dd\mathbbm{t} \quad ,
        \end{equation}
        which corresponds to squaring the eigenvalues.
\end{enumerate}
The architecture is illustrated in \reff{fig:Surmod}. In summary, the surrogate function $\surmod$ is composed of a neural network part, that yields a fourth order tensor with minor and major symmetries as well as orthorhombic symmetries, a subsequent filter layer to filter out anisotropic parts if required and finally a last layer to guarantee the positive semidefiniteness of the final output $\eltens$. With these modifications, $\surmod$ guarantees all imposed restrictions.
Once the parameters of $\surmod$ are adapted appropriately, inverse design can be performed with the following method.

\section{Inverse design}
\label{sec:inverse_design}
As $\surmod$ assumes alignment of the spinodoid with the coordinate axes, the function
\begin{equation}
    \label{eq:surmod_ext}
    \surmodQ(\strpar, \rottens) = \rottens\star\surmod(\strpar)
\end{equation}
is introduced, where $\rottens\star\eltens$ means $\bar C_{ijkl}Q_{mi}Q_{nj}Q_{ok}Q_{pl} \allowbreak \basisvec_m\dyad\basisvec_n\dyad\basisvec_o\dyad\basisvec_p$. The rotation tensor $\rottens\in\SOn{3}$ allows arbitrary 3d rotations of the spinodoid structure, which transforms the coordinates of the elasticity tensor according to \refe{eq:surmod_ext}. $\rottens$ is parametrized with three angles $\rotangles = (\varphi, \omega, \epsilon)\in \rotanglesspc = (0,\pi) \times (0,2\pi) \times (0,2\pi)$, where $\varphi$ and $\omega$ are the spherical coordinates of a point in the unit sphere specifying the axis of rotation $\tensorone{a}, |\tensorone{a}|=1$, and $\epsilon$ is the angle of rotation around this axis. The rotation tensor is assembled using Rodrigues' rotation formula
\begin{equation}
    \label{eq:rottens}
    \rottens(\rotangles) = \tensorone{a}\dyad\tensorone{a} + \cos(\epsilon)(\unittensor-\tensorone{a}\dyad\tensorone{a}) + \sin(\epsilon)\tensorone{a}\times\unittensor \quad .
\end{equation}
The two main advantages of a surrogate model $\surmodQ$ based on $\surmod$ as described above, are the short inference time of $\surmod(\strpar)$ compared to $\fwdmod(\strpar)$ as well as the analytically accessible derivatives with respect to $\strpar$.
These two advantages allow to solve inverse design problems very efficiently. In the following, these problems will be treated as minimization problems of the form
\begin{align}
    \label{eq:optprob}
    \strpar^*, \rotangles^* = \argmin_{\strpar\in\strparspc, \rotangles\in\rotanglesspc}\,&\objective(\strpar, \rottens(\rotangles)) \\
    \text{s.t.} \quad &g_p(\strpar, \rottens(\rotangles)) \leq 0, \quad p = 1, \dots, N_{\text{ieq}} \quad, \nonumber\\
    &h_q(\strpar,\rottens(\rotangles)) = 0, \quad q = 1, \dots, N_{\text{eq}} \quad , \nonumber
\end{align}
where the objective function $\objective$ encodes freely choosable target properties that are to be met, using the extended surrogate $\surmodQ$. Such objectives could include, for example, elastic moduli in specific directions, which can be derived from the output of $\surmodQ(\strpar, \rottens)=\eltens$.
The objective is to be minimized under problem specific inequality constraints $g_p$ and equality constraints $h_q$. Such constraints might be a requirement for a minimum stiffness $E_{\tensorone{d}}$ in a certain direction $\tensorone{d}$, i.e., $E_{\tensorone{d}}(\strpar, \rottens)\geq E_{\tensorone{d}}^{\text{min}}$ or a fixed density $\rho=0.5$.

Since $\strparspc$ is a non-connected domain, solving \refe{eq:optprob} as a whole is difficult in practice. Therefore, $\strparspc$ is divided into its seven non-connected subdomains. The space of allowed $\mangle_i$ is split into the regions $\manglespc=\manglespc_1\cup\manglespc_2$, where $\manglespc_1 = \{0\}$ and $\manglespc_2=(\SI{15}{\degree}, \SI{90}{\degree})$, resulting in the subdomains
\begin{align}
    \strparspc^{\text{lamel}}_1 &= \manglespc_2\times\manglespc_1\times\manglespc_1\times\rhospc \quad , \\
    \strparspc^{\text{lamel}}_2 &= \manglespc_1\times\manglespc_2\times\manglespc_1\times\rhospc \quad , \nonumber\\
    \strparspc^{\text{lamel}}_3 &= \manglespc_1\times\manglespc_1\times\manglespc_2\times\rhospc \quad , \nonumber\\
    \strparspc^{\text{colum}}_1 &= \manglespc_1\times\manglespc_2\times\manglespc_2\times\rhospc \quad , \nonumber\\
    \strparspc^{\text{colum}}_2 &= \manglespc_2\times\manglespc_1\times\manglespc_2\times\rhospc \quad , \nonumber\\
    \strparspc^{\text{colum}}_3 &= \manglespc_2\times\manglespc_2\times\manglespc_1\times\rhospc \quad , \quad \text{and} \nonumber\\
    \strparspc^{\text{cubic}} &= \manglespc_2\times\manglespc_2\times\manglespc_2\times\rhospc \quad . \nonumber
\end{align}
The minimization problem is solved within each of these subdomains, and the result with the smallest value for $\ell$ is selected. This approach is summarized in \refa{alg:ID}.
\begin{algorithm}[t]
    \caption{Piecewise solution of the optimization problem \refe{eq:optprob} by splitting $\strparspc$ into subregions.}
    \label{alg:ID}
    \begin{algorithmic}
        \Require $\ell, g_p, h_q$
        \State $\underline{\mathbf{S}}, \underline{\mathbf{Q}}, \underline{\boldsymbol\ell} = (), (), ()$
        \For{$\Tilde{\strparspc} \in \{ \strparspc^{\text{lamel}}_1, \strparspc^{\text{lamel}}_2, \ldots , \strparspc^{\text{cubic}} \}$}
            \State \vspace{-15pt}
                \begin{align*}
                    \hspace{0.5cm} \Tilde{\strpar}^*, \Tilde{\rottens}^* = \argmin_{\strpar\in\Tilde{\strparspc}, \rottens\in\SOn{3}}\,&\objective(\strpar, \rottens) \\
                    \text{s.t.} \quad &g_p(\strpar, \rottens) \leq 0\\
                    &h_q(\strpar,\rottens) = 0
                \end{align*}
            \State append $\Tilde{\strpar}^*$ to $\underline{\mathbf{S}}$, $\Tilde{\rottens}^*$ to $\underline{\mathbf{Q}}$, $\ell(\Tilde{\strpar}^*, \Tilde{\rottens}^*)$ to $\underline{\boldsymbol\ell}$
        \EndFor
        \State $i^*=\argmin_{i} \underline{\boldsymbol\ell}[i]$
        \State \Return $\underline{\mathbf{S}}[i^*], \, \underline{\mathbf{Q}}[i^*]$
    \end{algorithmic}
\end{algorithm}

For spinodoids, which have four structure parameters, \refe{eq:optprob} poses a bounded minimization problem with seven variables and equality/ inequality constraints. Solving the problem requires an appropriate optimizer. In this work, the SLSQP \cite{kraft1988software} algorithm is used, which is available through the Python package \texttt{scipy} \cite{2020SciPy-NMeth}.

\section{Data generation}
\label{sec:data_generation}
Calibrating the surrogate model $\surmod$ requires a dataset $\dataset=\{ (\strpar^\alpha, \eltens^\alpha) \}_{\alpha=1}^{\Ndata}$ in the form of pairs of structure parameters $\strpar^\alpha$ and corresponding stiffness tensors $\eltens^\alpha$. To create such a dataset, the structure parameters are sampled appropriately using an effective sampling strategy and the simulation pipeline $\fwdmod(\strpar)$ is evaluated for these sampled structure parameters to determine the corresponding stiffness tensors. Both steps are described below.

\subsection{Sampling}
\label{subsec:sampling}

The goal of the sampling is to generate a number $\Ndata$ of different tuples of structure parameters $\strpar$ for the training dataset to calibrate the parameters $\WeightsAndBiases$ of the network  as part of the surrogate $\surmod$. The number of required samples should be as small as possible while still covering the entire design space, so that the function $\fwdmod$ can be approximated adequately by $\surmod$. Since the architecture of the network guarantees equivariance with respect to permutations of $\mangle_1, \mangle_2, \mangle_3$, it is sufficient not to sample within the full domain $\strparspc$ but rather within a smaller region with $\mangle_1\geq\mangle_2\geq\mangle_3$, denoted as $\strparspctri\subset\strparspc$. As the design space is not connected (see \refe{eq:strpardomain}), all three cases with number of non-zero morphological angles $\Nnonzero=1,2,3$ (corresponds to lamellar, columnar and cubic structures) have to be considered. In order to sample the $\theta_i$ accordingly, a point $\matrixone{\xi}$ is first picked randomly within the hypercube domain $(0,1)^{\Nnonzero}$. This point $\matrixone{\xi}$ is then transformed into a region with $\vartheta_{j+1}\leq \vartheta_{j}$ using the transformation \footnote{Another option would be to simply trim the hypercube to a tetrahedral shape after sampling. However, this does not guarantee that exactly $\Ndata$ samples will remain in the desired domain.}
\begin{equation}
    \label{eq:tranformationTriangle}
    \vartheta_j = \sqrt[\Nnonzero-j+1]{\xi_j \vartheta_{j-1}^{\Nnonzero-j+1}} \quad \text{with} \quad \vartheta_0=1 \quad .
\end{equation}
Subsequently, the actual morphological angles $\theta_i$ result from scaling and shifting the domain to $\left\{ \SI{0}{\degree}\right\}\cup(\SI{15}{\degree},\SI{90}{\degree})$ via
\begin{equation}
    \label{eq:shiftdomaintheta}
    \mangle_i =   \begin{cases}
            \vartheta_i\cdot\SI{75}{\degree}+\SI{15}{\degree}\,,\quad \text{if } i\leq\Nnonzero \\
            \SI{0}{\degree}\hspace{1.75cm}, \quad \text{else} \quad .
        \end{cases}
\end{equation}
The volume fraction $\rho$ is obtained by sampling $\varrho$ within $(0,1)$ and transforming it into the admissible domain via $\rho=\varrho\cdot0.7+0.3$. 

This approach allows to generate structure parameters for cubic as well as columnar and lamellar spinodoids within $\strparspctri$. However, the purely random sampling of $\matrixone{\xi}$ and $\varrho$ has the disadvantage that clusters may arise and other regions of $\strparspctri$ are not scanned at all, which unnecessarily increases the dataset size without providing additional information for learning $\surmod$. Therefore, quasi-random Latin Hypercube Sampling is used instead, for both $\matrixone{\xi}$ and $\varrho$.

\begin{figure}
\centering
    \includegraphics{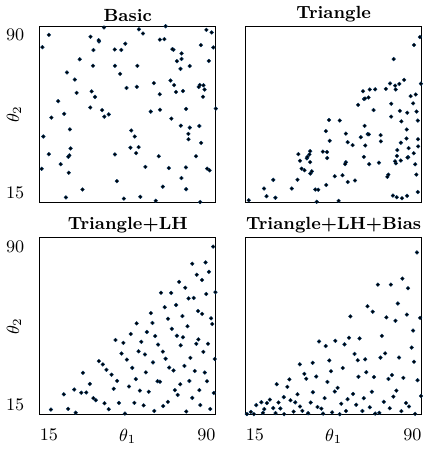}
    \caption{Sampling strategy of the $\mangle_i$ on the example of only $\mangle_1$ and $\mangle_2$: Purely random sampling in a quadratic and a triangular domain, Latin Hypercube (LH) sampling in the triangular domain and finally with a bias towards smaller values, as used for the training data sets.}
    \label{fig:Sampling}
\end{figure}

The sensitivity of the effective properties of spinodoid structures with small $\mangle_i$ and $\rho$ is very high, whereas changes at large $\mangle_i$, on the other hand, have a lower impact. Therefore, the sampling strategy should favor small $\theta_i$ and $\rho$. This is taken into account by introducing biases $\biastheta$ and $\biasrho$ as exponents, that shift $\matrixone{\vartheta}$ and $\varrho$ towards values closer to zero via $\Tilde{\vartheta}_j=\vartheta_j^{\biastheta}$ and $\Tilde{\varrho}=\varrho^{\biasrho}$ before transfoming them into their real parameter space. Note that $\vartheta_i, \varrho \in (0,1)$. The values of $\biastheta$ and $\biasrho$ are chosen to be $\biastheta=\biasrho=1.6$. \reff{fig:Sampling} shows the effect of sampling in the subspace, Latin Hypercube sampling and introducing a bias on the example of two non-zero morphological angles, i.e., columnar structures.

\subsection{Simulation framework}
For the sampled structure parameters $\strpar^\alpha$, respective geometries $\geo^\alpha$ are to be generated, and their effective elasticity tensors $\eltens^\alpha$ are to be determined. This workflow $\fwdmod(\strpar)$ includes the steps (i) generating a geometry $\geo^\alpha$ from given parameters $\strpar^\alpha$, (ii) assigning constitutive behavior to each phase, and (iii) carrying out homogenization simulations.

Firstly, geometries $\geo^\alpha$ are generated. These represent a finite section of the structure $\structure^\alpha$ associated with the structure parameters $\strpar^\alpha$. This section should be sufficiently large to reflect all the characteristics of the structure.
$\geo^\alpha$ is generated by evaluating \refe{eq:indicatorfunc} in a domain $\Omega$. As the sampling of the wave vectors $\wavevector_i$ and phase shifts $\phaseshift_i$ is randomized, it is advantageous to choose large $\Nwaves$ to minimize the impact of randomness, ideally with $\Nwaves\to\infty$. However, the computational costs for the evaluation of $\GRF(\tensorone{x})$ at a point $\tensorone{x}$ increases linearly with $\Nwaves$. We choose $\Nwaves=10,000$ within this work and generate cubic geometries in the domain $\Omega = (0,l)^3$ with $l=1$ and wave number $\wavenumber=30\pi$. The geometry is discretized with $\Nvoxel=128^3$ voxels. In each of the $\Nvoxel$ voxels, \refe{eq:indicatorfunc} is evaluated in the center of the voxel in order to assign either material $M_1$ or material $M_2$ to the voxel.

Both materials $M_i$ are modeled as linear elastic and isotropic with Young's modulus $E_i$ and Poisson's ratio $\nu_i$. Material $M_1$ is a stiff base material with normalized parameters $E_1=1$ and $\nu_1=0.3$ and material $M_2$ approximates the absence of any material with a Young's modulus 100 times smaller, i.e., $E_2=0.01$ and $\nu_2=0.3$. Note that not the absolute value of $E_1$ and $E_2$ is important but only their ratio, as the coordinates of $\eltens$ scale proportionally to $E_i$ if the ratio remains fixed.

The effective elasticity tensor of the voxelized two-phase geometry $\geo^\alpha$ with the materials $M_1$ and $M_2$ is determined using the FFT-based homogenization solver FANS (Fourier Accelerated Nodal Solvers \footnote{We use the open-source implementation available from \href{https://github.com/DataAnalyticsEngineering/FANS.git}{github.com/DataAnalyticsEngineering/FANS.git.}}) \cite{leuschner_fourier-accelerated_2018}.
To determine the effective elasticity tensor, six independent simulations are carried out, each with different prescribed effective strains ${}^{k}\bar\eps$ (here in Mandel notation)
\begin{equation}
    \mandel{{}^{1}\bar{\eps}} = \begin{bmatrix} \hat{\varepsilon}\\0\\0\\0\\0\\0 \end{bmatrix} , \,
    \mandel{{}^{2}\bar{\eps}} = \begin{bmatrix} 0\\\hat{\varepsilon}\\0\\0\\0\\0 \end{bmatrix} , \, \ldots \, , \,
    \mandel{{}^{6}\bar{\eps}} = \begin{bmatrix} 0\\0\\0\\0\\0\\\hat{\varepsilon} \end{bmatrix}
\end{equation}
with $\hat{\varepsilon}=\num{1e-6}$. With the obtained effective stresses ${}^{k}\bar{\sig}$, the entries of the stiffness tensor are obtained by solving the equations ${}^{k}\bar{\sig} = \eltens \dd {}^{k}\bar{\eps}$ using Mandel notation.

In this way, the corresponding elasticity tensor $\eltens^\alpha$ is assigned to each tuple of sampled structure parameters $\strpar^\alpha$ to obtain the structure-property pair $(\strpar^{\alpha}, \eltens^{\alpha})$. Using this workflow $\fwdmod$ consisting of parameter sampling and simulation framework, different data sets are generated to quantify the efficiency of the surrogate model $\surmod$ described in \refs{sec:surrogate_model}.

\subsection{Instances of data sets}
Training data sets of different sizes and a test data set are generated. The training data sets differ in the number of $(\strpar, \eltens)$-pairs and are generated using the method described in \refs{subsec:sampling}, i.e., with Latin Hypercube sampling in the modified design space $\strparspctri$ and with sampling bias $\biastheta=\biasrho=1.6$. The test dataset is generated with Latin Hypercube sampling as well, but covers the full design space without sampling bias.
The three types of spinodoids are present in equal parts in each data set. If $\Ndata$ is not divisible by 3, the cubic and, if necessary, the columnar spinodoids are preferred.
The parameters for the different data sets is summarized in \reft{tab:datasets}.
\begin{table}
    \caption{Generated datasets with relevant parameters.}
    \label{tab:datasets}
    \centering
    \begin{tabular}{cccc}
        \hline\noalign{\smallskip}
        Name & $\Ndata$ & $\biastheta$, $\biasrho$ & \makecell{Sampling\\ space}\\
        \noalign{\smallskip}\hline\noalign{\smallskip}
        $\dataset^{\text{test}}$ & 1000 & 1.0, 1.0 &  $\strparspc$\\
        $\dataset^{\text{train}}_{[\Ndata]}$ & \makecell{10, 20, 50, 75, 100,\\ 150, 200, 250,\\ 300, 500, 1000} & 1.6, 1.6 & $\strparspctri$\\
        \noalign{\smallskip}\hline
    \end{tabular}
\end{table}

\section{Applications}
\label{sec:applications}
The framework described so far is now being tested. First, the surrogate model $\surmod$ from \refs{sec:surrogate_model} is calibrated with datasets of different sizes, and the accuracy of the models are evaluated to find the minimal dataset size for accurate predictions. Subsequently, the chosed model is used to solve several inverse design tasks using the method described in \refs{sec:inverse_design}.
\subsection{Forward model}
\label{subsec:appl_forward_model}
\begin{figure}
    \includegraphics{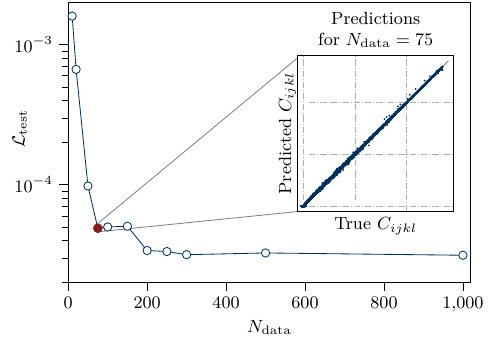}
    \caption{Error of surrogate predictions for the test data set. For every dataset size $\Ndata$, the model was trained 100 times with different initializations for the weights and biases and the best model was chosen based on the final error on the training data.}
    \label{fig:TDConvergence}
\end{figure}

We implement a surrogate model $\surmod$ as described in \refs{sec:surrogate_model} with hyperparameters summarized in \refs{sec:nnarchitecture}. The hyperparameters were chosen such that the number of trainable parameters is as small as possible while remaining flexible enough to adapt to the largest dataset $\dataset^{\text{train}}_{1000}$. 

With a given data set $\dataset$, the parameters are adapted according to
\begin{align}
    &\WeightsAndBiases^* = \argmin_{\WeightsAndBiases} \loss^{\text{data}} (\WeightsAndBiases; \dataset) + \lambda^{\text{reg}}\loss^{\text{reg}}(\WeightsAndBiases) \quad \text{with} \nonumber\\
    & \loss^{\text{data}} (\WeightsAndBiases; \dataset) = \frac{1}{n|\dataset|} \sum_{(\strpar, \hat{\eltens})\in\dataset} \norm{\surmod(\strpar; \WeightsAndBiases)-\hat{\eltens}}^2 \quad \text{, where} \nonumber\\
    & n = \max_\alpha \left( \|\hat{\eltens}^\alpha\|^2 \right) \quad \text{and} \nonumber\\
    &\loss^{\text{reg}}(\WeightsAndBiases) = \frac{1}{N^{\text{var}}}\sum_{i=1}^{N^{\text{var}}} w_i^2  \quad , \label{eq:nntraining}
\end{align}
where $N^{\text{var}}$ is the number of independent parameters (weights and biases) of the network and $w_i$ is the $i$-th of those independent parameters and $\lambda^{\text{reg}}=\num{1e-4}$. For each dataset the minimization problem \refe{eq:nntraining} is solved 100 times with different random initializations of the weights and biases. We use SLSQP for the training process. For each dataset size, the model with the lowest final value of the loss function is kept, all others are discarded. Those ten chosen model are tested with the unseen dataset $\dataset^{\text{test}}$ with the values of the loss function $\loss(\WeightsAndBiases; \dataset^{\text{test}})$ shown in \reff{fig:TDConvergence}. The figure shows that training datasets with more than 200 data points do not further improve accuracy. Even the model trained with 75 data points is sufficiently accurate. The corresponding correlation plot is also included in \reff{fig:TDConvergence}. All other test results are presented in \refs{sec:prediction}.

Here, we choose the model trained with $\dataset^{\text{train}}_{75}$ and use it for the subsequent inverse design tasks. Of course, both $\dataset^{\text{train}}_{75}$ and $\dataset^{\text{test}}$ were used in this study when selecting the model. Nevertheless, it becomes evident that the required number of data points is significantly lower than used in previous works, which can be explained by the low complexity of the function to approximate, see \refs{sec:sensitivity}. Our findings are particularly relevant for more expensive properties, such as for inelastic behavior, and enables estimates for the required number of training data for other problems.

\subsection{Inverse design}
\label{subsec:appl_inverse_design}
To validate the inverse design framework, three examples of varying complexity are presented. In the first example, a full, realistic stiffness tensor is prescribed and a suitable spinodoid structure is to be predicted by the inverse design approach. The subsequent examples focus on optimizing Young's moduli in specific directions while simultaneously minimizing the volume fraction of the base material. In order to shorten notation, the function \cite{nordmann_visualising_2018}
\begin{equation}
    E(\mathbbm{C}, \tensorone{d}) = \frac{1}{(\tensorone{d}\dyad\tensorone{d})\dd\mathbbm{C}^{-1}\dd(\tensorone{d}\dyad\tensorone{d})}
\end{equation}
is introduced for calculating the Young's modulus of a material with elasticity tensor $\mathbbm{C}$ with respect to a specific direction $\tensorone{d}$, where $\norm{\tensorone{d}}=1$.

\paragraph{Example 1}
\begin{figure}
    \includegraphics{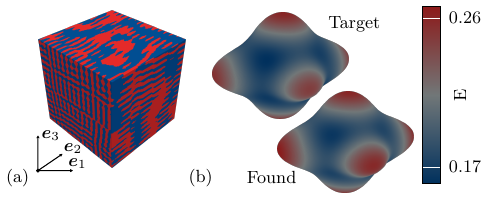}
    \caption{Result of inverse design Example 1, reconstructing a full target elasticity tensor: (a) the identified geometry and (b) elastic surfaces of the target and the found geometry.}
    \label{fig:ID01}
\end{figure}
As a minimal example for validating the framework, a complete elasticity tensor is prescribed, and the corresponding structure parameters are to be found. An elasticity tensor not present in the training or test dataset is used as target value, generated with the simulation pipeline $\fwdmod$ using the structure parameters $\hat\strpar = (\SI{20}{\degree}, \SI{20}{\degree}, \SI{20}{\degree}, 0.5)$, i.e., $\hat{\eltens}=\fwdmod(\hat\strpar)$. The objective function for solving the inverse design problem utilizes the squared norm of the deviation of network prediction from the target and is given by
\begin{equation}
    \ell(\strpar, \rottens) = \frac{1}{\|\hat{\eltens}\|}\norm{\hat{\eltens} - \surmodQ(\strpar, \rottens)} \quad .
\end{equation}
There are no constraints , meaning $N_{\text{ieq}}=N_{\text{eq}}=0$.
Solving \refe{eq:optprob} yields structure parameters $\strpar^*=(\SI{19.4}{\degree}, \allowbreak \SI{20.6}{\degree}, \SI{20.0}{\degree}, 0.506)$ and rotation tensor $\rottens^*$ with Rodrigues angles $\rotangles^*=(\SI{0.3}{\degree}, \SI{45.2}{\degree}, \SI{181.4}{\degree})$, corresponding to roughly a half rotation around $\basisvec_3$. The final value of the objective function is $\ell(\strpar^*, \rottens^*)=\num{1.9e-4}$.
Comparing $\hat\strpar$ and $\strpar^*$ suggests sufficient agreement of $\eltens^*=\fwdmod(\strpar^*)$ with the desired $\hat{\eltens}$. Nevertheless, the proposed structure is homogenized for comparison, yielding a stiffness tensor $\eltens^*$ with $||\hat{\eltens}-\eltens^*|| / ||\hat{\eltens}|| = \SI{4.4}{\percent}$. Note that the generation of spinodoid structures as described in \refs{sec:spinodoids} is stochastic, resulting in slightly different effective properties for different instances. The good compliance becomes evident in the nearly indistinguishable elastic surfaces in \reff{fig:ID01}.

\paragraph{Example 2}
\label{par:Ex02}
\begin{figure}
    \includegraphics{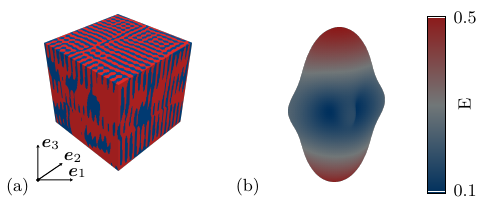}
    \caption{Result of inverse design Example 2, finding a spinodoid with lowest possible volume fraction $\rho$ and Young's modulus in $\basisvec_3$-direction larger than $\bar E_{\basisvec_3}^{\text{min}}=0.5$: (a) the identified geometry and (b) its elastic surface.}
    \label{fig:ID02}
\end{figure}
The goal of this inverse design problem is to find the spinodoid structure that has an effective Young's modulus in the $\basisvec_1$-direction $\bar E_{\basisvec_1}$ greater than $\bar E_{\basisvec_1}^{\text{min}}=0.5$ with the minimal possible volume fraction $\rho$. The minimization of the volume fraction $\rho\rightarrow\min$ is achieved through the objective function
\begin{equation}
    \ell(\strpar, \rottens) = \rho^2 \quad ,
\end{equation}
while the minimum allowed Young’s modulus in $\basisvec_1$-direction is implemented as an inequality constraint
\begin{align}
    \label{eq:ID02Constr}
    &g_1(\strpar, \rottens) = \bar E_{\basisvec_1}^{\text{min}} -\bar E_{\basisvec_1} \leq 0 \quad\quad,\quad\text{where} \\
    &\bar E_{\basisvec_1} = E\left(\surmodQ(\strpar, \rottens), \basisvec_1\right) \nonumber
\end{align}
Solving the optimization problem \refe{eq:optprob} yields $\strpar^*=(\SI{15}{\degree}, \SI{15}{\degree}, \SI{0}{\degree}, 0.547)$ and $\rotangles^*=(\SI{3}{\degree}, \SI{0}{\degree}, \SI{168}{\degree})$. The corresponding structure is shown in \reff{fig:ID02} and has a Young’s modulus in the $\basisvec_1$-direction of $\bar E_{\basisvec_1}^*=0.503$. Modifying the found structure parameters results in metamaterials with worse behavior: a slight reduction in the volume fraction to $\rho=0.53$ results in a structure with $\bar E_{\basisvec_1}=0.476$ falling below the desired value, whereas small changes in $\mangle_1$, $\mangle_2$ and $\mangle_3$ also lead to a violation of condition $g_1$. Thus, the identified structure at least describes a local minimum. This framework is therefore also capable of solving inverse problems of this type.

\paragraph{Example 3}
A spinodoid structure is to be found that simultaneously meets the following three conditions:
\begin{enumerate}[label=(\roman*), leftmargin=*]
    \item The effective Young’s modulus in the $\tensorone{d}_1$-direction must be greater than the required minimum value $\bar E_{\tensorone{d}_1}^{\text{min}}=0.3$,
    \item the ratio $q=\bar E_{\tensorone{d}_2}/\bar E_{\tensorone{d}_3}$ of the effective Young’s moduli in $\tensorone{d}_2$-direction and $\tensorone{d}_3$-direction must be close to $\hat{q}=2$, and
    \item the amount of required base material should be minimized, i.e., $\rho\rightarrow\min$.
\end{enumerate}
\begin{figure}[t]
    \includegraphics{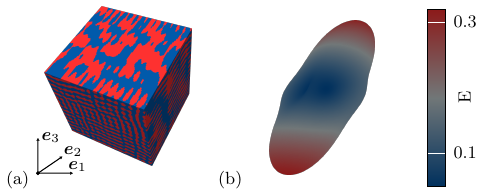}
    \caption{Result of inverse design Example 3, finding a spinodoid with lowest possible volume fraction $\rho$, Young's modulus in a direction $\tensorone{d}_1$-direction larger than $\bar E_{\tensorone{d}_1}^{\text{min}}=0.3$ and ratio of Young's moduli in two other perpendicular directions of $\hat{q}=2$: (a) the identified geometry and (b) its elastic surface.}
    \label{fig:ID03}
\end{figure}
The preferred directions $\tensorone{d}_i=\hat{\rottens}\cdot\basisvec_i$ are prescribed with $\hat{\rottens}$ using \refe{eq:rottens} and the Rodrigues angles $\hat{\rotangles}=(\SI{20,1}{\degree},\SI{71,0}{\degree},\SI{180}{\degree})$, which corresponds to a rotation of $\basisvec_1$ towards $\tensorone{d}_1=(\frac{1}{\sqrt{3}},\frac{1}{\sqrt{3}},\frac{1}{\sqrt{3}})$ via the rotation axis $(\basisvec_1+\tensorone{d}_1/) |\basisvec_1+\tensorone{d}_1|$.
For this inverse design problem, the objective function
\begin{align}
    \label{eq:ID03Obj}
    &\ell(\strpar, \rottens) = \ell^{q}(\strpar, \rottens) + \ell^{\rho}(\strpar, \rottens) \quad, \text{where} \\
    &\ell^{q}(\strpar, \rottens) = \left( \frac{\bar E_{\tensorone{d}_2}}{\bar E_{\tensorone{d}_3}} - \hat{q} \right)^2 / \hat{q}^2 \quad \text{and} \nonumber\\
    &\ell^{\rho}(\strpar, \rottens) = \rho^2 \nonumber
\end{align}
is formulated, which is to be minimized under the inequality constraint
\begin{equation}
    \label{eq:ID03Constr}
    g_1(\strpar, \rottens) = \bar E_{\tensorone{d}_1}^{\text{min}} - \bar E_{\tensorone{d}_1} \leq 0 \
\end{equation}
already known from Example 2. In Eqs. (\ref{eq:ID03Obj}) and (\ref{eq:ID03Constr}), the directional Young's moduli are calculated through $\bar E_{\tensorone{d}_i}= E\left(\surmodQ(\strpar, \rottens), \tensorone{d}_i\right)$.
With \refe{eq:optprob}, the structure parameters $\strpar^*=(\SI{15}{\degree}, \SI{0}{\degree}, \SI{22.8}{\degree}, 0.431)$ and Rodrigues angles $\rotangles^*=(\SI{45}{\degree}, \SI{159}{\degree}, \SI{227}{\degree})$ are obtained. These parameters result in $\ell^{q}=\num{1.3e-9}$ and $\ell^{\rho}=0.284$. The structure realized with these parameters has a Young’s modulus of $\bar E_{\tensorone{d}_1}^*=0.319$, satisfying condition (i). Additionally, condition (ii) is well met with $\frac{\bar E_{\tensorone{d}_2}^*}{\bar E_{\tensorone{d}_3}^*}=2.09$.
Other sets of structure parameters with small variations in $\mangle_i$, $\rho$ and $\rotangles$ do not result in a spinodoid metamaterial, that satisfies all of the conditions. Thus, this inverse design problem could be solved as well.

\section{Summary}
\label{sec:summary}
We showed that precise inverse design of complex targeted properties using spinodoid metamaterials is possible with a very small dataset of structure parameter-stiffness pairs. Previous approaches mostly focus on reconstructing complete stiffness tensors or individual effective Young's moduli, requiring several thousand data points or considering only parts of the full design space. Our approach enables the creation of a precise, physically sound surrogate with only 75 data points and allows for optimizing arbitrary objectives using the presented inverse design framework.

Spinodoid metamaterials offer advantageous mechanical properties and, despite their low-dimensional description, exhibit a rich property space, making them particularly interesting for inverse design problems. The four parameters include the volume fraction of the base material $\rho$ and three morphological angles $\theta_i$. Permutations of these angles result in structures that are equivalent in a specific way, which is also reflected in their effective properties.
We exploit this feature to develop a data-efficient surrogate model based on a permutation-equivariant neural network. This network architecture enforces this property exactly by construction rather than learning it from data.
Using this special architecture, a surrogate model is created to map structure parameters to the effective elasticity tensor. In addition to permutation equivariance, it strictly satisfies several other essential properties.

Herein, inverse design is interpreted as a minimization problem with respect to the structure parameters and 3d rotations. The presence of a surrogate model enables efficient solving of the minimization problem and the incorporation of arbitrary constraints. This allows to tackle arbitrary inverse design tasks, ensuring broad applicability of the framework.

The proposed method is then validated with regard to multiple aspects. To train the model, datasets of varying sizes are generated, where the design space is sampled appropriately according to observed phenomena like equivariance with respect to permutations of the morphological angles and higher sensitivity with smaller structure parameters. The model is calibrated using these datasets. Its predictive accuracy is validated against a large test dataset. The results show that 75 data points are sufficient to obtain a model with adequate precision.
Using this model, multiple inverse design tasks of increasing complexity are solved. The suggested solutions are evaluated for their plausibility. The framework successfully solves all tasks.

This demonstrates that the proposed framework is more efficient and flexible compared to other approaches. While data efficiency is of secondary importance for the linear-elastic properties studied here --since simulations with the FANS software used are cost-effective -- this insight is highly relevant for more advanced inverse design tasks involving complex mechanical behavior, such as nonlinear elasticity under large deformations or inelastic properties like a viscoelastic stress response. Additionally, it makes the creation of a purely experimental dataset for surrogate model calibration much more feasible.

The demonstrated data reduction shows that creating precise surrogate models for the mapping between spinodoid structure parameters and effective properties is well possible even for more complex properties without requiring unfeasible datasets, which in turn enables the solution of arbitrary inverse design problems.

\paragraph{Acknowledgements}
This work was supported by the German Research Foundation (DFG) within the Research Training Group GRK 2868 ``Data-Driven Design of Resilient Metamaterials'' (D$^3$) -- project number 493401063.
Most presented computations were performed on a HPC Cluster at the Center for Information Services and High Performance Computing (ZIH) of TU Dresden. The authors thank the ZIH for the generous allocations of computer time. The authors also thank Alexandra Otto for providing the code for geometry and mesh generation as well as Karl A. Kalina and Jörg Brummund for the fruitful discussions during the development of the work.

\paragraph{CRediT authorship contribution statement}
\textbf{Max Rosenkranz}: Conceptualization, Methodology, Formal Analysis, Software, Investigation, Validation, Visualization, Writing - original draft, Writing - review \& editing. \textbf{Markus Kästner:} Conceptualization, Funding acquisition, Supervision, Writing - review \& editing. \textbf{Ivo F. Sbalzarini:} Conceptualization, Funding acquisition, Supervision, Writing - review \& editing.

\appendix

\section{Network architecture}
\label{sec:nnarchitecture}
The permutation equivariant network, that is used in the surrogate model as described in \refs{sec:surrogate_model}, maps the four structure parameters $\strpar$ to a fourth order tensor with minor and major symmetry and zeros on all the indices that correspond to zeros in an orthorhombic tensor of fourth order with preferred directions aligned with the coordinate axes. The input is decomposed as a 3-vector for $(\mangle_1, \mangle_2, \mangle_3)$ and a scalar for $\rho$. The scalar $\rho$ is treated as a matrix of order zero, which means, that it is connected to every output orbit of the first layer via an independent weight.
The network has two hidden layers with ten neurons of rank one each and softplus activation $sp$, which is defined as
\begin{equation}
    sp(x) = \ln(1+e^x) \quad .
\end{equation}
The output layer has a single neuron of rank four with linear activation. With the restrictions explained in \refs{sec:surrogate_model}, the full network comprises 313 free parameters.
Input and output are normalized to range $(-1,1)$ within each orbit using the respective training data set. Note that normalizing each input/output coordinate independently would violate permutation equivariance.

\section{Prediction accuracy}
\label{sec:prediction}
\begin{figure*}
    \def\svgwidth{\linewidth}
    \centering
    \includegraphics[width=\textwidth]{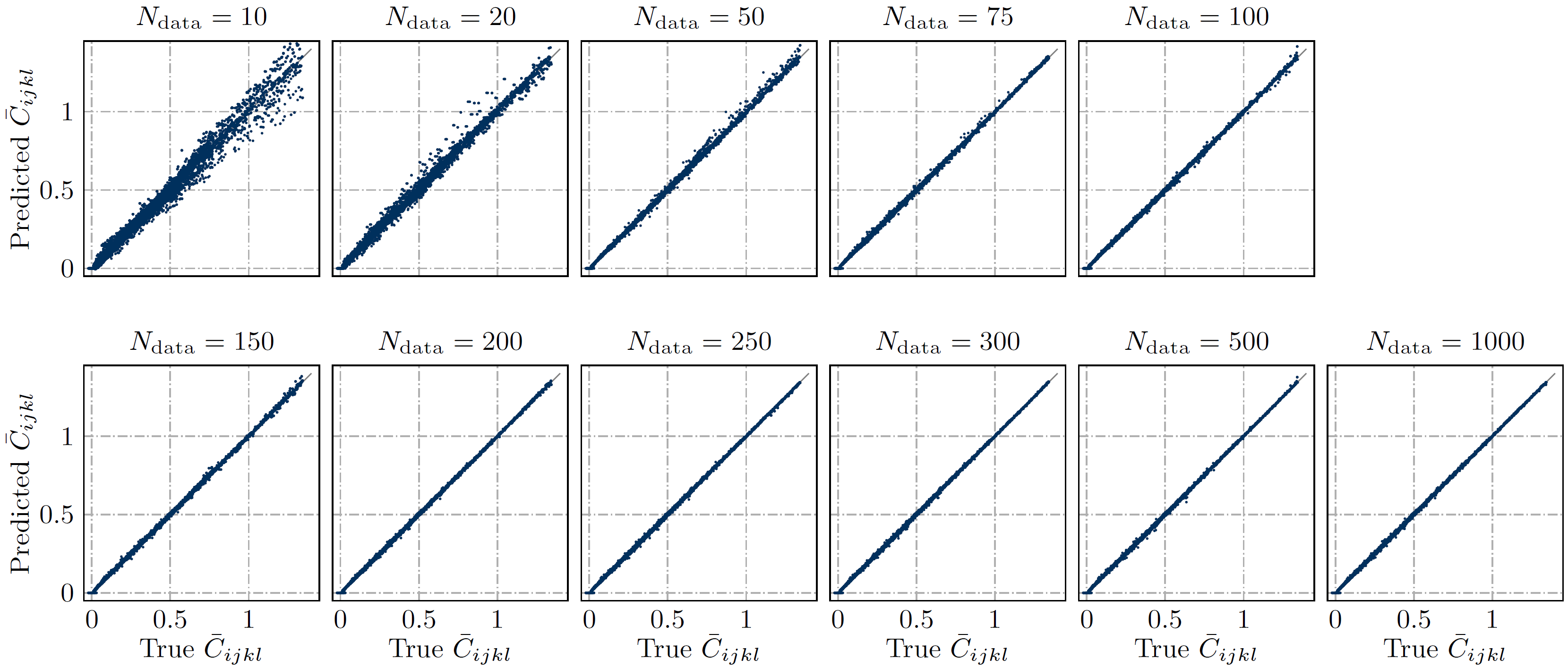}
    \caption{Overview of the surrogate model's prediction quality for different training dataset size $\Ndata$. For each $\Ndata$, the model was trained 100 times, and the model with the smallest final training loss was selected. It becomes evident that even small datasets are entirely sufficient to gather sufficient information for the whole design space.}
    \label{fig:PredictionsAll}
\end{figure*}

In \refs{subsec:appl_forward_model}, the model trained with only 75 data points was chosen for the later inverse design applications. This decision is based on \reff{fig:TDConvergence}, which shows the error on the test dataset $\loss^{\text{test}}$ of the best model (based on $\loss^{\text{train}}$) plotted against the number of training data points. Since this value $\loss^{\text{test}}$ is not particularly intuitive, the corresponding correlation plots are provided in \reff{fig:PredictionsAll}. It becomes clear that using more than 100 data points offers only negligible improvements in accuracy. The choice of the model with $\Ndata=75$ remains somewhat arbitrary, however, \reff{fig:PredictionsAll} is intended to justify this decision.

\section{Sensitivity of the properties with respect to the structure parameters}
\label{sec:sensitivity}
\begin{figure*}
    \centering
    \includegraphics{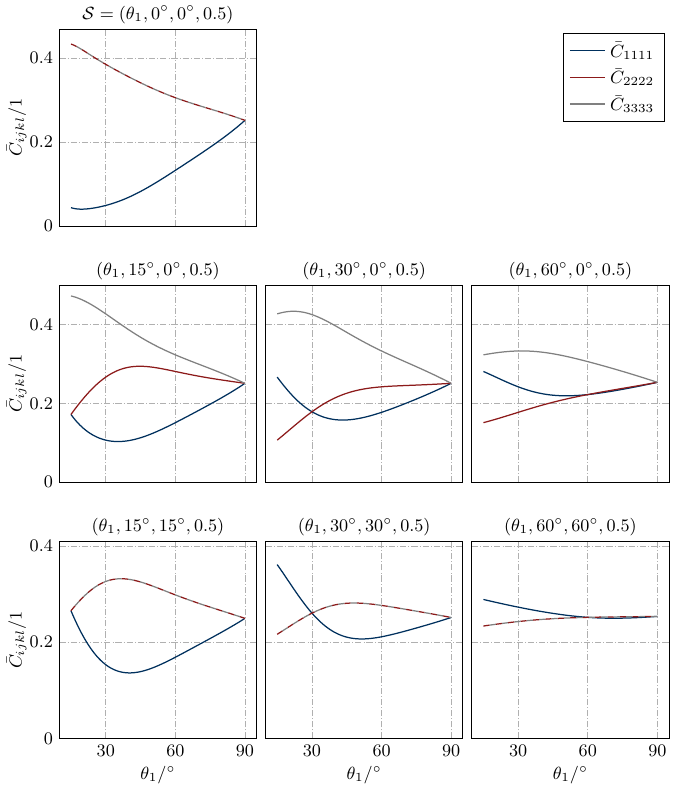}
    \caption{Complexity of the function $\fwdmod: \strpar \mapsto \eltens$, evaluated for several cross sections of the design space $\strparspc$ using a surrogate model that was trained on 1,000 data points. The rather smooth and non-oscillating nature of the curves demonstrates the low complexity of the function the surrogate model has to learn.}
    \label{fig:sensitivity}
\end{figure*}

In order to investigate the complexity of the function $\fwdmod$ that is to be learnt by the model $\surmod$, the model calibrated with 1,000 data points is evaluated for various combinations of structure parameters. This model has sufficient accuracy to carry out the following analyses without having to perform simulations for each parameter combination, see \reff{fig:PredictionsAll}.
The model is evaluated for varying $\mangle_1$ and various fixed values for $\mangle_2, \mangle_3$ and $\rho$. 
The corresponding curves of the coordinates $C_{1111}, C_{2222}$ and $C_{3333}$ of the stiffness tensor $\eltens=\surmod(\strpar)$ are plotted in \reff{fig:sensitivity}. The graphs shows smooth curves without oscillations and no more than one local extremum. The function to be learnt by the surrogate model is therefore of fairly low complexity, which explains the low need for data.

\bibliographystyle{unsrtnat}

\end{document}